\documentstyle[12pt]{article}

\begin{document}

\newcommand{\be}{\begin{equation}}
\newcommand{\ee}{\end{equation}}
\newcommand{\ba}{\begin{eqnarray}}
\newcommand{\ea}{\end{eqnarray}}
\newcommand{\f}{\frac}
\newcommand{\s}{\sqrt}

\begin{center}
{\bf ON THE PHASE OF THE OFF-SHELL SCATTERING AMPLITUDES}\\ 
\vskip \baselineskip
{\bf V.A.Petrov}\\

\vspace*{1mm}
{\it Division of Theoretical Physics, IHEP, 
Protvino, Russia}\\
\end{center}

\vspace*{5mm}
\begin{abstract}
We prove that the phase of the partial wave amplitude does not depend
on the
off-shell 4-momentum squared of one of the scattered particles. 
\end{abstract}

\vspace*{10mm}
Let $T_l(s)$ is the partial wave amplitude for elastic scattering with
angular
momentum $l$ and c.m.s. energy $\sqrt s$. Assuming that there is no
unphysical
thresholds we consider ``elastic region'' in $s$, where unitarity
takes a
simple
form
\be
Im\;T_l(s)=\left |T_l(s)\right |^2.
\ee
Let now one of the four interacting particles is off-shell, and denote
the
corresponding amplitude by $T_{l*}\equiv T_l(s,q^2\not= m^2)$. If two
particles
(one final and one initial)  are off mass-shell then $T_{l**}$ is the
corresponding amplitude. For simplicity we take all off-shell momenta
below the
mass-shell. 

In this case for the same ``elastic region'' one can get from
unitarity that 
\be
Im\,T_{l**}=\left |T_{l*}\right |^2,
\ee
\be
ImT_{l*}=ReT_lRe T_{l*}+ Im T_l Im T_{l*}. 
\ee
In terms of the following parametrisation
$$
T_l=r_le^{i\varphi_l},\;\;
T_{l*}=r_{l*}e^{i\varphi_l*},\;\;
T_{l**}=r_{l**}e^{i\varphi_l**}\;\;
$$
we have from Eqs.(1), (2) and (3)
$$
\sin\varphi_l=r_l,
$$
$$
r_{**l}\sin\varphi_{l**}=r^2_{l*},
$$
$$
\sin\varphi_{l*}= \sin^2\varphi_{l} \sin\varphi_{l*}
+\sin\varphi_{l} \cos\varphi_{l} \cos\varphi_{l*}, 
$$
whereof we came to the equality: 
$$
\varphi_{l*}=\varphi_{l}.
$$

That is the main result of this note:\\

{\bf onefold off-shell extension of the partial wave scattering
amplitude in the ``elastic region'' in the c.m.s. energy does not
change the
phase.}
In particular 
$$
T_{l}(s,q^2\not=m^2)=r_{l} (s,q^2\not=m^2)T_l(s),
$$
where $r_{l} (s,q^2)$ is a real-analytic function in $s$ and has
no elastic threshold singularity.

We also have to stress that for the ``full'' amplitude, 
\ba
T(s,\cos\Theta,q^2) &\left |\right.&  = \f{16\pi\s s}
{\s {s-4m^2}}\sum_{l}
(2l+1)T_{l*}P_l(\cos\Theta),  \nonumber \\[-0.4mm]
 q^2 &\not =& m^2 
\ea
its phase generally depends on $q^2$-value. 

It is a pleasure to thank A.Martin and T.T.Wu for useful discussions.

\enddocument